%% file: acl_latex.tex
\pdfoutput=1

\documentclass[11pt]{article}
\usepackage{multirow}

\usepackage[final]{acl}

\usepackage{times}
\usepackage{latexsym}

\usepackage[T1]{fontenc}

\usepackage[utf8]{inputenc}

\usepackage{microtype}

\usepackage{inconsolata}

\usepackage{graphicx}

\title{GitGoodBench: A Novel Benchmark For Evaluating Agentic Performance On Git}

\author{
\textbf{Tobias Lindenbauer\textsuperscript{1,2}}
\thanks{Work done during an internship at JetBrains},
\textbf{Egor Bogomolov\textsuperscript{1}},
\textbf{Yaroslav Zharov\textsuperscript{1}}
\\
\textsuperscript{1}JetBrains Research\\
\textsuperscript{2}School of Computation, Information and Technology, Technical University of Munich
\\
\small{
\textbf{Correspondence:} \href{mailto:tobias.lindenbauer@jetbrains.com}{tobias.lindenbauer@jetbrains.com}
  }
}

\usepackage[nolist,nohyperlinks]{acronym}
\usepackage{subcaption}
\usepackage{amsmath}
\usepackage[frozencache]{minted}
\usepackage{booktabs}
\usepackage{hyperref}
\usepackage{tcolorbox}
\usepackage{cleveref}
\input{acronym}
\begin{document}

\maketitle
\begin{abstract}
Benchmarks for \ac{swe} AI agents, most notably SWE-bench, have catalyzed progress in programming capabilities of AI agents. However, they overlook critical developer workflows such as \ac{vcs} operations. To address this issue, we present GitGoodBench\footnote{\url{https://github.com/JetBrains-Research/git-good-bench}}, a novel benchmark for evaluating AI agent performance on \acf{vcs} tasks. GitGoodBench covers three core Git scenarios extracted from permissive open-source Python, Java, and Kotlin repositories. Our benchmark provides three datasets: a comprehensive evaluation suite (900 samples), a rapid prototyping version (120 samples), and a training corpus (17,469 samples). We establish baseline performance on the prototyping version of our benchmark using GPT-4o equipped with custom tools, achieving a 21.11\% solve rate overall. We expect GitGoodBench to serve as a crucial stepping stone toward truly comprehensive \ac{swe} agents that go beyond mere programming.
\end{abstract}

\begin{figure}[htbp]
    \centering
    \begin{subfigure}[b]{0.41\textwidth}
        \centering
        \includegraphics[width=\textwidth]{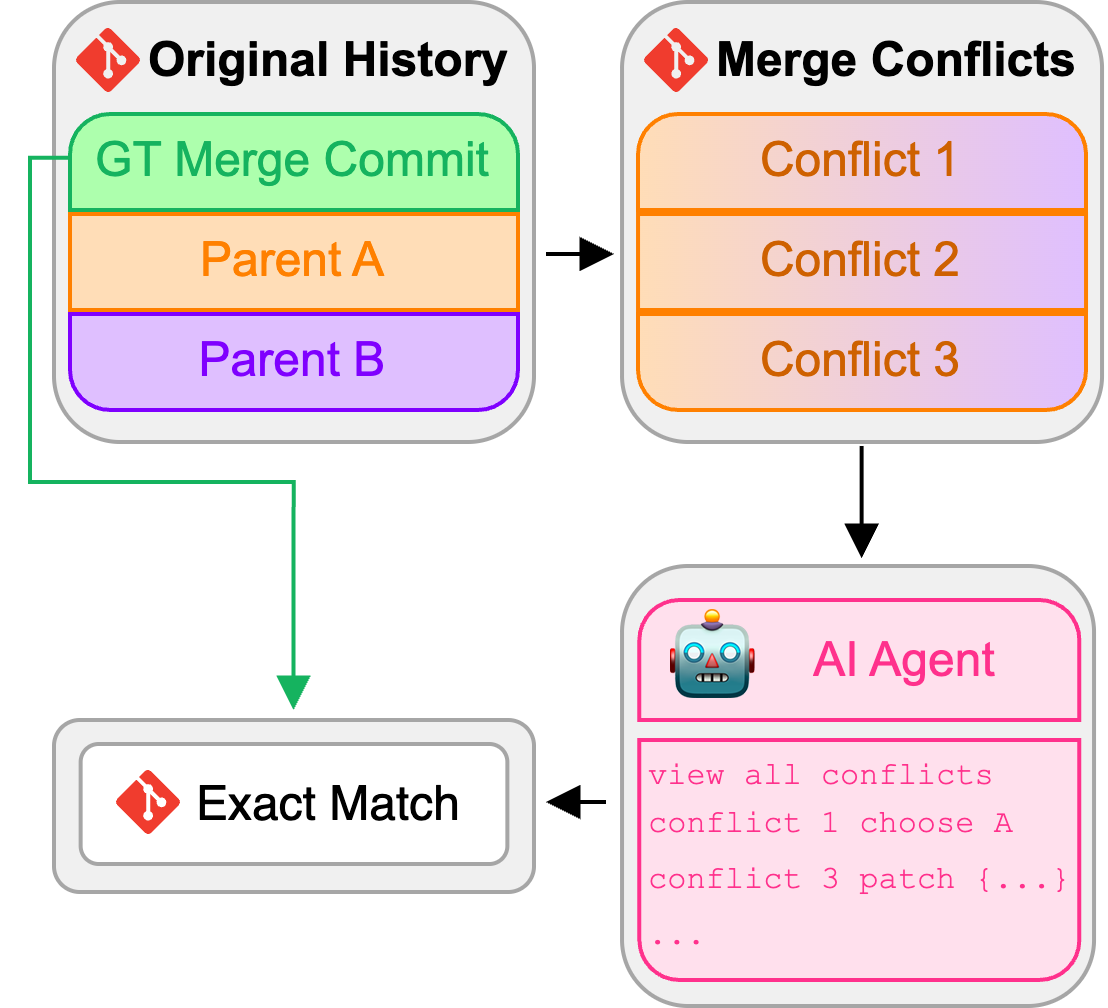}
        \caption{Merge Conflict Resolution: The agent must reproduce the ground-truth merge commit given a set of conflicts.}
        \label{fig:mcr}
        
    \end{subfigure}
    \hfill 
    \begin{subfigure}[b]{0.41\textwidth}
        \centering
        \includegraphics[width=\textwidth]{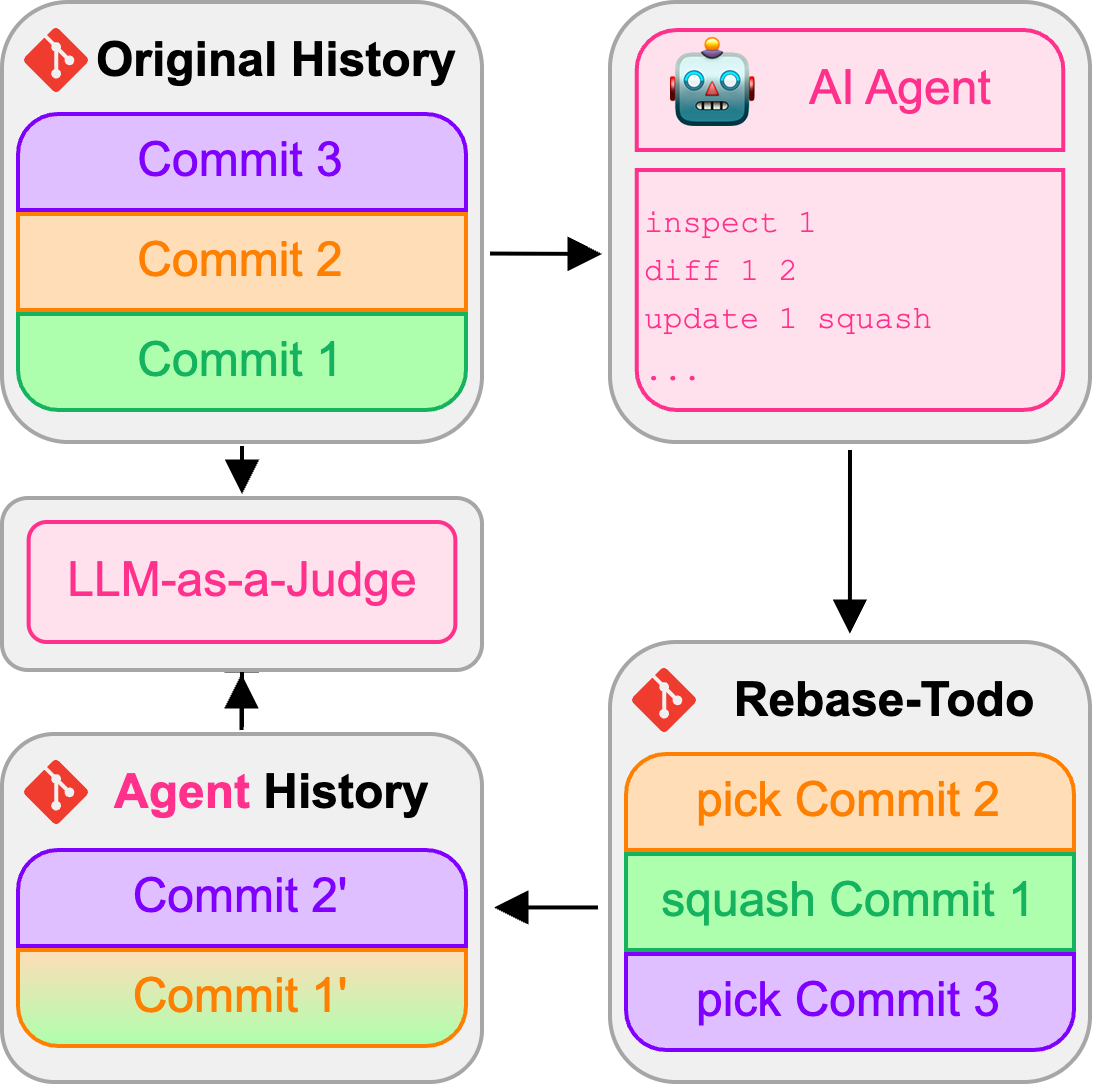}
        \caption{Interactive Rebase: The agent generates an alternative history based on existing commits.}
        \label{fig:ir}
    \end{subfigure}
    \hfill 
    \begin{subfigure}[b]{0.41\textwidth}
        \centering
        \includegraphics[width=\textwidth]{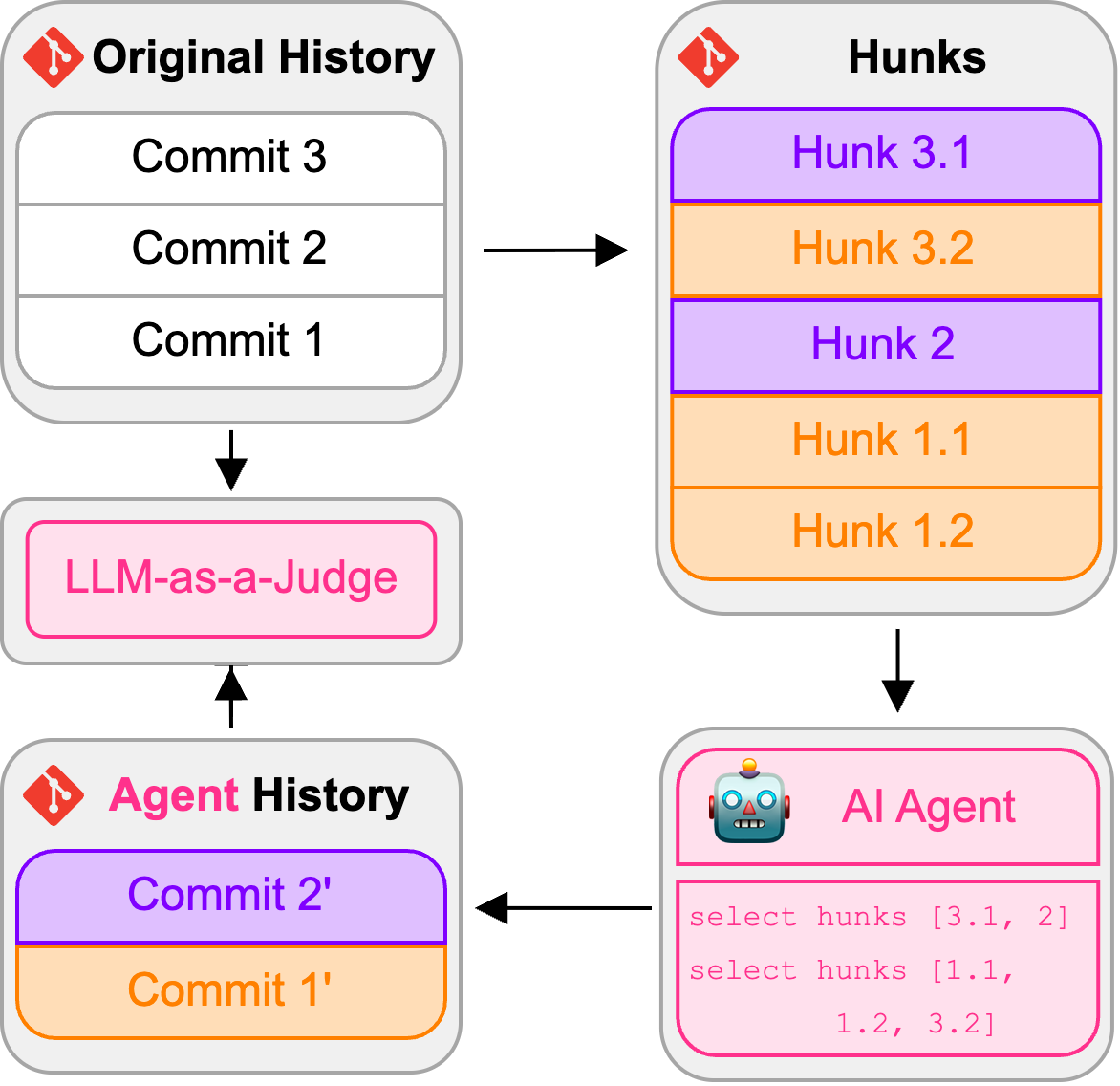}
        \caption{Iterative Committing of Changes: The agent generates an alternative based on a disorganized set of changes. We only use the original commit history for evaluation.}
        \label{fig:icc}
    \end{subfigure}
    
    \caption{The three Git scenarios supported by GitGoodBench. Each scenario benchmarks a typical Git use-case and unique aspect of version control.}
    \label{fig:git-workflows}
\end{figure}

\section{Introduction}
While the rapid scaling of~\acp{llm} has led to promising results across various tasks initially, the improvements gained from scaling models further are slowing down. Compared to GPT-3~\cite{brown_language_2020}, GPT-3.5 achieves a approximately $60\%$ improvement~\cite{openai_gpt-4_2024} on MMLU~\cite{hendrycks_measuring_2021}. The improvement from GPT-3.5 to GPT-4, however, is just approximately $23\%$~\cite{openai_gpt-4_2024}. Scaling test-time compute rather than just models has emerged as an alternative for further improving performance, leading to the rise of AI agents~\cite{yao_react_2023,shinn_reflexion_2023,wang_executable_2024}. AI agents equip~\acp{llm} with external tools~\cite{schick_toolformer_2023} and employ sophisticated planning and reasoning strategies such as ReAct~\cite{yao_react_2023} or Reflexion~\cite{shinn_reflexion_2023} to dynamically adjust in uncertain environments. 

\acf{swe} emerged as a pivotal application domain due to the availability of high-quality data in open-source repositories and because the creation and maintenance of software underpins innovation and economic impact across virtually every sector. SWE-bench~\cite{jimenez2024swebench} is the industry-standard benchmark for evaluating the agent's programming proficiency through testing the agent's ability to fix bugs in real-world software. This spurred the rapid development of AI agents for programming by major players in the tech tooling ecosystem~\cite{cursor-agent,google-agent,jetbrains-agent,vscode-agent,claude-sonnet-and-code-agent}. 

\acfp{vcs}, such as Git, are ubiquitous in \ac{swe}~\cite{cortes_rios_unifying_2022} and play a pivotal role in building software in distributed teams. It is thus natural to use Git as a medium of collaboration between AI agents and human engineers. While \ac{llm} providers are advertising the Git capabilities of their systems~\cite{claude-sonnet-and-code-agent}, there currently exists no benchmark for evaluating an AI agent's capacity of interacting with Git in an end-to-end manner. Furthermore, typical Git tasks such as \acf{ir} are time-consuming and distinct from raw code-generation. \ac{ir} requires reasoning over the Git history and an in-depth understanding of dependencies between the commits constituting the history.

To stimulate innovation in the direction of comprehensive, end-to-end \ac{swe} AI agents that go beyond mere programming, we introduce a novel benchmark for the popular \ac{vcs} Git. This comprises a training corpus for collecting agentic trajectories and two evaluation sets (lite and full). The benchmark supports \acf{mcr}, \acf{ir}, and the \acf{icc} (\Cref{fig:git-workflows}). We scrape all data from permissive, open-source, Python, Java, or Kotlin GitHub repositories. Furthermore, we provide a baseline implementation using GPT-4o~\cite{openai-gpt-4o} with custom tools, achieving a 21.11\% solve rate.

\section{Related Work}
Several benchmarks, such as SWE-bench~\cite{jimenez2024swebench}, or the Kowinski prize~\cite{konwinski-prize} evaluate agentic systems on complex, multi-turn \ac{swe} tasks sourced from real-world GitHub issues. While the environment allows Git usage,  the evaluation focuses solely on whether the agent resolves the bug rather than how it leverages VCS. In contrast, our benchmark explicitly measures an agent’s proficiency with Git tasks. This allows future research to thoroughly examine and refine \ac{vcs}-focused strategies in \ac{swe} agents and tailor agents to \ac{vcs} tasks specifically.

While previous works on automating or evaluating \ac{mcr}~\cite{svyatkovskiy_program_2022,shen_git_2023,boll_towards_2024,pan_can_2021} and commit message generation or completion~\cite{jiang_automatically_2017,hal_generating_2019,eliseeva_commit_2023} exist, they exclusively cater to specific \ac{vcs} subtasks. In contrast, our benchmark is the first to encapsulate multiple subtasks, such as commit message generation, reasoning across commits, and rebase plan generation into a single benchmarking scenario. This uniquely positions GitGoodBench for evaluating and training AI agents with expertise in \ac{vcs} tasks in end-to-end settings.

\begin{table*}[ht]
\centering
\begin{subtable}{\textwidth}  
\small
\centering
\begin{tabular}{ll}
        \hline
        \textbf{Filter} & \textbf{Reason} \\ \hline
        Repository size $\leq 400$MB & Avoid Git LFS heavy repositories \\ 
        Repository stars $\geq 1000$ & Heuristic for repository quality \\ 
        Repository is not archived & Heuristic for repository quality \\ 
        Repository is not forked & Avoid duplication \\ 
        Last commit within a month of May, 31st 2024 & Avoid stale repositories \\ 
        Repository has permissive license & Ensure legal compliance \\ 
        Repository $\geq 5$ branches & Heuristic for merge conflict scenarios \\ 
        Repository $\geq 5$ contributors & Heuristic for merge conflict scenarios \\ 
        \hline
    \end{tabular}
    \caption{Repository metadata filters we use for selecting the initial repositories we consider in the benchmark creation. We consider the following licenses permissive: MIT, Apache 2.0, BSD 3-Clause ``New'' or ``Revised'', BSD 2-Clause ``Simplified''.}
    \label{tab:filters:metadata}
\end{subtable}
\vfill
\vspace{.25cm}
\begin{subtable}{\textwidth}
\small
\centering
\begin{tabular}{ll}
        \hline
        \textbf{Filter} & \textbf{Reason} \\ \hline
        No merge commit in \ac{fcc} & Cleanly separate scenario types\\
        No merge conflict in unsupported file type & Only support Python, Java, or Kotlin \\
        All merge scenarios contain conflict & Merges without a conflict are trivial \\
        Merge scenarios have exactly two parents & Avoid dilution by complex and rare merge types \\
        Number of merge conflicts $\leq 8$ & Ensure the agent can theoretically solve the scenario \\
        Length of \ac{fcc} $\leq 6$ & Ensure the agent can theoretically solve the scenario \\
        \ac{fcc} file is modified, not added & Otherwise we get a single hunk when resetting \\
        \hline
    \end{tabular}
    \caption{Scenario level filters for selecting scenarios to include in our benchmark.}
    \label{tab:filters:scenario}
    \end{subtable}
    \label{tab:filters}
    \vspace{-20pt}
    \caption{Filters for selecting repositories and scenarios to include in our benchmark.}
\end{table*}

\section{GitGoodBench Datasets}
We provide \href{https://huggingface.co/datasets/JetBrains/git_good_bench}{GitGoodBench} (900 samples) and \href{https://huggingface.co/datasets/JetBrains/git_good_bench-lite}{GitGoodBench Lite} (120 samples) for evaluation in comprehensive and rapid-prototyping settings, respectively. The research community recently started investigating \ac{swe} agents powered by fine-tuned \acp{slm}~\cite{pan_swegym_2024,jain_r2e-gym,yang_swe-smith_2025}. We believe that trained, on-device sized agents are an exciting research direction. While we do not train such a model in this work, with \href{https://huggingface.co/datasets/JetBrains/git_good_bench-train}{GitGoodBench Train} (17,469 samples) we release a dataset split dedicated to collecting trajectories for training Git agents.

\subsection{Supported Scenarios}
\label{subsec:dataset:supported-scenarios}
Our benchmark covers the following three types of Git scenarios:
\begin{description}
\item[\acl{mcr}] The agent must resolve all merge conflicts by reproducing the ground truth resolutions (\Cref{fig:mcr}). 

\item[\acl{ir}] In this scenario (\Cref{fig:ir}) the agent must reason across commits and their contents to determine the optimal ordering of commits, thereby improving the Git history. This includes commit consolidation or modification and commit message refinement.

\item[\acl{icc}] This scenario (\Cref{fig:icc}) type is the inverse of the \ac{ir}. Instead of optimizing existing commits, the agent must generate a reasonable Git history from a large disorganized set of changes.
\end{description}

With these scenario types we cover non-trivial Git functionalities central to common Git workflows~\cite{cortes_rios_unifying_2022}. Moreover, we explicitly cover functionality currently only implemented interactively in Git (e.g., \texttt{git rebase -i} or \texttt{git add -p}). Agents are highly applicable for such iterative tasks that depend on environment observations. However, interacting with such functionality is challenging for agentic systems because these functions do not provide immediate feedback and instead wait for user input. This introduces friction into the typical plan-act-observe loop of AI agents, due to delayed feedback not easily captured by usual pipelines.

\subsection{Dataset Creation}
\label{subsec:dataset:creation}
We collect repository metadata from repositories with permissive licenses using SEART~\cite{Dabic:msr2021data} and the metadata filters defined in \Cref{tab:filters:metadata}.  The scenarios for \ac{ir} and \ac{icc} are represented by the same samples in our dataset (i.e., with one sample, we can evaluate both \ac{ir} and \ac{icc}). We call these samples \acf{fcc} samples and they refer to chains of commits in Git histories in which we observe consecutive modifications of a single file. We use this as a heuristic for identifying Git histories that may be improved through reordering or consolidating commits. 
These samples target the use-case of (1) cleaning up the local Git history before pushing new commits to the remote (e.g., \texttt{git rebase -i HEAD\textasciitilde5}, and (2) constructing a clean git history given a set of changes for the \ac{ir} and \ac{icc} scenario, respectively. To tailor these samples toward evaluating an aspect of Git distinct from \ac{mcr}, we remove merge commits from \acp{fcc}. This allows us to evaluate the system's understanding of the rebase-todo and of relationships between commits.
We then mine the Git history of these repositories for merge, and \ac{fcc} samples and apply our scenario-level filters (\Cref{tab:filters:scenario}) to obtain 6,917 merge samples and 11,572 \ac{fcc} samples. To ensure a diverse benchmark, especially concerning represented repositories, we partition our data into strata based on the following features before sampling to construct our benchmark. 

\paragraph{\acl{fcc} Samples} For these samples, we use the project size (in lines of code) and the repository name for stratification.

\paragraph{\acl{mcr} Samples} In addition to the above, we stratify on the difficulty of these samples. We define \ac{mcr} difficulty based on the number of conflicts and their distribution across files. To determine conflicts, we run \texttt{git show --remerge-diff <merge-commit>} and identify conflicts through Git merge conflict markers. We consider scenarios with a single conflict ``easy'' because no reasoning across diffs is necessary, those with multiple conflicts in a single file ``medium'' because reasoning across diffs in the context of a single file is required, and all others, for which the agent must reason across multiple diffs and files, as ``hard''. 

To construct the held-out test, we sample 120 scenarios for GitGoodBench Lite and 900 for GitGoodBench. We stratify the sampling for scenario type and \ac{pl}. The remaining samples yield GitGoodBench Train. All three datasets are mutually disjoint. For further details, see \Cref{sec:dataset-appendix}.

\subsection{Metrics}
\label{subsec:dataset:metric}
We present the results of our baseline in terms of success and solve rate (both expressed as percentages). The \emph{success rate} refers to scenarios for which our system did not cause an error (e.g., because a patch cannot be applied in \ac{mcr}). Below, we define the \emph{solve rate} for each scenario:

\paragraph{\acl{fcc} Samples} For \ac{fcc} scenarios we prompt an \ac{llm} to judge the agent-generated and ground truth Git histories using the LLM-as-a-Judge~\cite{zheng_judging_2023} approach. We opt for this approach instead of \ac{em}, because there is no clear, deterministic way to define what constitutes a superior Git history. Following~\citet{zheng_judging_2023} we judge each pair of Git histories twice while switching the positions of the histories in the same prompt template to account for position bias. We prompt the judge to base its decision on (1) the quality of the commit messages considering the contents of the commit, (2) the cohesion of changes within the commits, (3) a logical progression of changes across commits, and (4) the size of commits. If the judge chooses the agent-generated over the ground truth Git history in both cases, we count a sample as solved. For details on the prompt see \Cref{appendix:subsection:llm-eval}.

\paragraph{\acl{mcr} Samples} Because an exact ground truth solution is available, we use \ac{em} between the ground truth solution and the agent's solution for evaluating \ac{mcr}.

\section{Environment}
\label{sec:experiment}
As a baseline, we evaluate GPT-4o~\cite{openai-gpt-4o} on GitGoodBench Lite and the tasks defined in~\Cref{subsec:env:tasks} using the metrics in~\Cref{subsec:dataset:metric}. While we do not use an agentic reasoning framework~\cite{yao_react_2023,shinn_reflexion_2023,wang_executable_2024}, we do equip the \ac{llm} with one possible set of custom tools (\Cref{subsec:env:tools}). 

\begin{table}[t]
\centering
\begin{tabular}{l|c|c}
\hline
\textbf{Scenario} & \textbf{Success Rate} & \textbf{Solve Rate} \\
\hline
\ac{ir} & 93.33 & 26.67 \\
\ac{icc} & 93.33 & 23.33 \\
\ac{mcr} & 76.67 & 13.33 \\
\hline
\textbf{Total} & 88 & 21.11 \\
\hline
\end{tabular}
\caption{Success and solve rates (\%) by scenario type, rounded to two decimal places. We observe the high complexity of the proposed benchmark, even given the strong baseline model and custom environment tools.}
\label{tab:success-rates-scenarios}
\end{table}

\begin{table}[t]
\centering
\begin{tabular}{l|c|c}
\hline
\textbf{Difficulty Level} & \textbf{Success Rate} & \textbf{Solve Rate} \\
\hline
Easy & 80.64 & 22.58 \\
Medium & 84.62 & 7.69 \\
Hard & 62.5 & 0 \\
\hline
\end{tabular}
\caption{Success and solve rates (\%) by difficulty for \ac{mcr} samples, rounded to two decimal places. GitGoodBench Lite contains 31 ($\approx 52\%$) easy, 13 ($\approx 22\%$) medium, and 16 ($\approx 27\%$) hard samples.}
\label{tab:success-rates-difficulty}
\end{table}

\subsection{Provided Context}
\label{subsec:env:tasks}
\paragraph{\acl{ir}} In the initial context, we provide all changes in all commits participating in the \ac{ir}, few-shot function-calling examples and an explanation of valid commands for the rebase-todo file. We initiate the \ac{ir} covering all commits in the \ac{fcc} before launching the agent.

\paragraph{\acl{icc}} We provide all Git-generated hunks that the agent must process, in addition to few-shot function-calling examples in the initial context. After each commit, we automatically show the agent the updated list of remaining hunks. We limit the agent's selection of hunks to hunks originating from the file for which we mined the \ac{fcc} and commit all other changes in a single commit after the agent terminates.

\paragraph{\acl{mcr}} The initial context includes the temporal ordering of the commits being merged, names of all files with conflicts and all merge conflicts it must resolve as well as few-shot function-calling examples.

\subsection{Provided Tools}
\label{subsec:env:tools}
Initially we experimented with minimalistic tooling, simply giving the \ac{llm} terminal access in a sandbox environment. However, preliminary results indicated that the system is unable to make any meaningful progress in this setup\footnote{We acknowledge that a Git \ac{mcp} may address this issue but as the focus of our work is a benchmark, we do not further investigate this.}. In particular it struggled with interactive Git functionality (\Cref{subsec:dataset:supported-scenarios}. Because of this we opt for the strong scaffolding detailed below.

\paragraph{\acl{ir}} We implement tools for viewing the contents of commits and interacting with the rebase-todo list, a file that specifies how Git should carry out the \ac{ir}.

\paragraph{\acl{icc}} With our tooling for this scenario type, the agent selects any number of Git-generated hunks to group into a single commit.

\paragraph{\acl{mcr}} To foster coherent, conflict-spanning resolutions, we provide tools for viewing individual merge conflicts, complete files or the overall difference between commits being merged. Our tooling limits the agent to sequentially resolving conflicts. It may only specify a patch for resolving the current conflict.

\section{Baseline Results}
\label{sec:results}
In \Cref{tab:success-rates-scenarios}, we see that our baseline implementation succeeds in 88\% and solves 21.11\% of scenarios in GitGoodBench Lite\footnote{We release the raw evaluation data with our \href{https://github.com/JetBrains-Research/git-good-bench}{repository}.} overall. Even with significant scaffolding support the~\ac{llm} is unable to solve the majority of tasks in our benchmark. This highlights the need to explicitly consider Git use-cases when engineering and training \ac{swe} agents. 

For both \ac{ir} and \ac{icc} scenarios our system achieves higher success and solve rates than 
for \ac{mcr} scenarios (\Cref{tab:success-rates-scenarios}). We partially attribute to the stricter scaffolding for these two scenarios. In \ac{mcr} scenarios the agent must generate code that can be applied at the location of the conflict to solve the conflict. Especially in scenarios which require the agent to make globally consistent conflict resolution choices (i.e., medium and hard samples in \Cref{tab:success-rates-difficulty}) the system's performance rapidly deteriorates. In \ac{fcc}-based scenarios, the agent must simply select a set of hunks to commit for \ac{icc} scenarios or modify the rebase-todo file through a tool for \ac{ir} scenarios. This indicates that the failure rate of agentic systems interacting with Git increases as the level of technical abstraction from Git decreases. We do however note that some amount of this performance degradation may also be due to the stricter \ac{em} evaluation metric used for \ac{mcr} scenarios. Regarding the difficulty heuristic for \ac{mcr}, we note that it accurately captures a sample's complexity regarding the solve rate. Easy samples have a $\approx3$ times higher solve rate than hard samples. Furthermore, the scenarios based on \ac{fcc} samples (\ac{ir} and \ac{icc}) result in similar success and solve rates. This indicates that our LLM-as-a-Judge evaluation methodology is consistent in assessing similar Git histories and is thus a suitable choice. Our difficulty heuristic for \ac{ir} and \ac{icc} scenarios did not correlate with the observed difficulty, for details see \Cref{subsec:appendix:purity}.

\section{Conclusions}
GitGoodBench is a novel benchmark for training and evaluating AI agents on the Git scenarios: \ac{mcr}, \ac{ir} and \ac{icc}. Our baseline implementation demonstrates capabilities in resolving merge conflicts and improving Git histories when equipping GPT-4o~\cite{openai-gpt-4o} with tools for interacting with Git, achieving an overall solve rate of $21.11\%$ on GitGoodBench Lite. The poor overall performance and the observed performance degradation for \ac{mcr} across difficulty levels highlight the need to explicitly consider Git when designing \ac{swe} agents. Just as we construct agents for \ac{swe} with repository-level reasoning and code generation in mind, we should consider the agents' understanding of Git artifacts and capacity to use Git functionality. We hope our benchmark spurs innovation in this direction.

\section{Limitations}
Our baseline implementation has several constraints that present opportunities for improvement. The \ac{mcr} tooling cannot modify Git-generated hunk boundaries, limiting flexibility when these hunks are too coarse. For \ac{icc}, expanding beyond a single-file focus would allow more accurate handling of multi-file changes. Furthermore, enabling commit content modification during \ac{ir} would allow handling more complex \ac{ir} scenarios, including ones during which a merge conflict occurs. Additionally, for \ac{fcc} samples our evaluation methodology may introduce bias, as it is \ac{llm}-based. We suggest that future work evaluating agents on GitGoodBench use an ensemble of \acp{llm} for judging trajectories to mitigate bias and subjectivity of the evaluation. Finally, we did not investigate how a Git implementation of the novel \acf{mcp}~\cite{anthropic-mcp} affects an agent's ability to solve Git tasks.

Regarding the dataset itself, while we made efforts to ensure diversity, certain limitations remain. While our difficulty heuristic for \ac{mcr} showed promising results, a \ac{fcc} difficulty heuristic based on \ac{fcc} purity (\Cref{subsec:appendix:purity}) didn't correlate with empiric performance. Due to this, the distribution of \ac{fcc} samples may be skewed with respect to their difficulty in our benchmark. While our three scenario types cover core Git functionality, our benchmark does not yet include important Git diagnostic workflows such as \texttt{git bisect}. Incorporating bisect scenarios would enable evaluation of an AI agents' ability to systematically locate commits introducing bugs, a capability that could significantly enhance automated debugging and regression analysis in \ac{swe} AI agents. Furthermore, as our benchmark is static, we may need to update our benchmark with more diverse and complex scenarios to counteract benchmark saturation and data leakage. 

\section*{Acknowledgements}
We thank Yury Khudyakov, Alexandra Eliseeva, Maria Tigina, and Abhiram Bellur for the valuable discussions and advice during this project.

\bibliography{custom}

\appendix
\label{sec:appendix}
\begin{table*}[t]
\centering
\begin{tabular}{l||ccc|c||ccc|c}
\hline
\textbf{Scenario Type} & Easy & Medium & Hard & \textbf{Success Rate} & Easy & Medium & Hard & \textbf{Solve Rate} \\
\hline
\ac{ir} & 100 & 86.36 & 95.52 & 93.33 & 13.33 & 31.82 & 30.43 & 26.67 \\
\ac{icc} & 100 & 90.91 & 91.3 & 93.33 & 20 & 27.27 & 21.74 & 23.33 \\
\ac{mcr} & 80.64 & 84.62 & 62.5 & 76.76 & 22.58 & 7.69 & 0 & 13.33 \\
\hline
\textbf{Total} & 90.16 & 87.72 & 85.48 & 88 & 19.67 & 24.56 & 19.35 & 21.11 \\
\hline
\end{tabular}
\caption{Success and solve rates (\%) by scenario type and difficulty, rounded to two decimal places. GitGoodBench Lite contains 31 easy, 13 medium, 16 hard \ac{mcr} samples and 15 easy, 22 medium, and 23 hard \ac{fcc} samples.}
\label{tab:appendix:total-success-rates}
\end{table*}

\section{Dataset Details}
\label{sec:dataset-appendix}
In this section we provide further details about the diversity of our datasets with respect to represented repositories, and \ac{mcr} difficulty. For GitGoodBench Train we also provide information on the distribution across programming languages, for all other datasets this distribution is fixed to ensure diversity (see \Cref{subsec:dataset:creation}). Please also refer to our dataset cards on HuggingFace: \href{https://huggingface.co/datasets/JetBrains/git_good_bench-lite}{GitGoodBench Lite}, \href{https://huggingface.co/datasets/JetBrains/git_good_bench}{GitGoodBench}, \href{https://huggingface.co/datasets/JetBrains/git_good_bench-train}{GitGoodBench Train}.

In \Cref{tab:datasets-repository-diversity} we provide statistics on the diversity of our datasets with respect to the repositories represented. Notably, there is a heavy skew toward Python and to a lesser extent Java. However, this is in line with our expectations and the popularity of the programming languages we consider in our datasets. \Cref{tab:difficulty_distribution} provides further information regarding the distribution of \ac{mcr} difficulties across our datasets. We note that the difficulty of \ac{mcr} is overall relatively well-distributed with a spike in difficulty on GitGoodBench. Despite stratifying on difficulty, these spikes can occur because we also stratify on other features such as the programming language.
\begin{table*}[ht]
\centering
\label{tab:repo_stats}
\begin{tabular}{l|ccc}
\toprule
\textbf{Statistic} & \textbf{GitGoodBench Lite} & \textbf{GitGoodBench} & \textbf{GitGoodBench Train} \\
\midrule
Total Repositories & 100 & 479 & 816 \\
Mean Samples Per Repo & 1.20 & 1.87 & 21.40 \\
Standard Deviation & 0.79 & 2.80 & 48.80 \\
Minimum & 1 & 1 & 1 \\
25th Percentile & 1 & 1 & 2 \\
Median & 1 & 1 & 6 \\
75th Percentile & 1 & 2 & 18 \\
Maximum & 8 & 46 & 644 \\
\bottomrule
\end{tabular}
\caption{The diversity of our datasets with respect to unique repositories from which we mined our samples. Our datasets consist of 816 (525 Python, 284 Java, and 79 Kotlin) unique repositories overall. 
}
\label{tab:datasets-repository-diversity}
\end{table*}

\begin{table*}[ht]
\centering
\begin{tabular}{lccc}
\toprule
\multirow{2}{*}{\textbf{Difficulty}} & \textbf{GitGoodBench} & \textbf{GitGoodBench} & \textbf{GitGoodBench} \\
 & \textbf{Lite} & & \textbf{Train} \\
\midrule
Easy & 51.67 & 41.33 & 51.65 \\
Medium & 21.67 & 24.44 & 18.39 \\
Hard & 26.67 & 34.22 & 29.97 \\
\bottomrule
\end{tabular}
\caption{Difficulty distribution (in \%) across GitGoodBench datasets.}
\label{tab:difficulty_distribution}
\end{table*}

\subsection{Sample Data}
\Cref{tab:sample_structure} shows the complete structure of a data point in our dataset. The detailed contents of the \texttt{scenario} field vary depending on the \texttt{sample\_type} and are presented in \Cref{sec:scenario-field}.

\begin{table*}[ht]
\centering
\small
\begin{tabular}{p{0.2\textwidth}p{0.25\textwidth}p{0.45\textwidth}}
\toprule
\textbf{Field} & \textbf{Value} & \textbf{Description} \\
\midrule
id & \texttt{mockito\_mockito\_merge\_0002} & Unique identifier \\
\midrule
name & \texttt{mockito/mockito} & Repository name (owner/repository) \\
\midrule
default\_branch & \texttt{main} & Primary repository branch \\
\midrule
license & \texttt{MIT License} & Repository license \\
\midrule
stargazers & 14,617 & GitHub stars count \\
\midrule
created\_at & \texttt{2012-10-13T08:27:12} & Repository creation date \\
\midrule
topics & \texttt{java;java-library;mock;...} & Repository topics/tags \\
\midrule
programming\_language & \texttt{java} & Primary language \\
\midrule
scenario & \texttt{<scenario-details>} & Scenario-specific data (see \Cref{tab:sample_scenario_mcr,tab:sample_scenario_fcc}) \\
\midrule
sample\_type & \texttt{merge} & Type of code sample \\
\midrule
project\_size & \texttt{medium} & Estimated project size \\
\midrule
difficulty & \texttt{easy} & Complexity level \\
\bottomrule
\end{tabular}
\caption{Structure of a sample data point from our dataset. Each entry contains metadata about the repository, along with scenario-specific information that varies based on the sample type. The \texttt{topics} field is truncated for brevity.}
\label{tab:sample_structure}
\end{table*}

\begin{table*}[!t]
\centering
\small
\begin{tabular}{ll}
\toprule
\textbf{Field} & \textbf{Description} \\
\midrule
\texttt{file} & The relative path of the file this sample refers to. \\
\midrule
\texttt{branch} & The branch name from which this \ac{fcc} originates. \\
\midrule
\texttt{times\_seen\_consecutively} & The number of times this particular file was modified in succession. \\
\midrule
\texttt{purity} & $\in [0;1]$. Ratio between changes in the \texttt{file} and the total changes in all files \\ 
& of a \ac{fcc} scenario. \\
\midrule
\texttt{newest\_commit} & The commit hash corresponding to the newest or last commit in this \ac{fcc}. \\
\midrule
\texttt{oldest\_commit} & The commit hash corresponding to the oldest or first commit in this \ac{fcc}. \\
\midrule
\texttt{contains\_non\_pl\_files} & A boolean indicating whether any commit in this sample includes changes to\\
&files with types not covered by the supported \acp{pl}. \\
\bottomrule
\end{tabular}
\caption{Contents For \ac{fcc} Samples. \Cref{tab:sample_scenario_fcc} shows a representative example of the scenario field from our dataset. Due to a \texttt{purity} of $0.68$, we consider this sample to be of medium difficulty. We define the purity-based difficulty we investigated in more detail in \Cref{subsec:appendix:purity}.}
\label{tab:fcc_samples_contents}
\end{table*}

\subsection{The Scenario Field}
\label{sec:scenario-field}
In this section we provide further details regarding the contents of the \texttt{scenario} for the two sample types in our datasets.

\begin{table*}[t]
\centering
\small
\begin{tabular}{ll}
\toprule
\textbf{Field} & \textbf{Description} \\
\midrule
\texttt{merge\_commit\_hash} & The ground truth merge commit in which the conflicts are resolved. \\
\midrule
\texttt{parents} & List of parent commit hashes of the merge commit. \\
\midrule
\texttt{number\_of\_files\_with\_merge\_conflict} & The overall number of distinct files in which a merge conflict occurs. \\
\midrule
\texttt{total\_number\_of\_merge\_conflicts} & Total number of distinct merge conflicts across all files. \\
\midrule
\texttt{files\_in\_merge\_conflict} & Relative paths of the files that contain merge conflicts. \\
\bottomrule
\end{tabular}
\caption{Contents For Merge Conflict Resolution (MCR) Samples.}
\label{tab:mcr_samples_contents}
\end{table*}

\begin{table*}[t]
\centering
\small
\begin{tabular}{ll}
\toprule
\textbf{Field} & \textbf{Value} \\
\midrule
merge\_commit\_hash & \texttt{baa37f65fdff5b780a50d5b5c6bf8bc3ade43815} \\
\midrule
parents & \texttt{[d758810c59a9134f437d60f73a82036749688ccb,} \\ 
& \texttt{~5dcd493c67ff863c69c1214f0892a80e4951087e]} \\
\midrule
number\_of\_files\_with\_merge\_conflict & 2 \\
\midrule
total\_number\_of\_merge\_conflicts & 2 \\
\midrule
files\_in\_merge\_conflict & \texttt{[cogs/gpt\_3\_commands\_and\_converser.py,} \\
& \texttt{~models/openai\_model.py]} \\
\bottomrule
\end{tabular}
\caption{A sample \acf{mcr} scenario field from GitGoodBench Lite. Each entry contains metadata about a specific merge conflict instance, including commit identifiers and statistics about the conflicting files.}
\label{tab:sample_scenario_mcr}
\end{table*}

\subsubsection{Contents For \ac{fcc} Samples}
In \Cref{tab:fcc_samples_contents} we show the structure of the scenario field for \ac{fcc} samples. Furthermore, \Cref{tab:sample_scenario_fcc} provides an exemplary \ac{fcc} datapoint's \texttt{scenario} field contents. The \texttt{scenario} contains information regarding the source of the sample (e.g., the branch from which it was mined), the length of the \ac{fcc} and its starting and end commits.

\begin{table*}[t]
\centering
\small
\begin{tabular}{p{0.28\textwidth}p{0.62\textwidth}}
\toprule
\textbf{Field} & \textbf{Value} \\
\midrule
file & \texttt{composer/models/huggingface.py} \\
\midrule
branch & \texttt{origin/vincent-mlflow-logger-verbose} \\
\midrule
times\_seen\_consecutively & 3 \\
\midrule
purity & 0.68 \\
\midrule
newest\_commit & \texttt{c24b29f19c4c131a3ea7098dd8b8a5edde344819} \\
\midrule
oldest\_commit & \texttt{c1ff80900f46d4e36feb4b326689fe14fc41cbc6} \\
\bottomrule
\end{tabular}
\caption{A sample \acf{fcc} scenario field from GitGoodBench Lite. This example records a file's modification pattern across multiple commits, including branch information and a purity metric defined in \Cref{subsec:appendix:purity} and \Cref{subsec:dataset:supported-scenarios}.}
\label{tab:sample_scenario_fcc}
\end{table*}

\subsubsection{Contents For Merge Samples}
In \Cref{tab:fcc_samples_contents} we detail the structure of the scenario field for \ac{mcr} samples. \Cref{tab:sample_scenario_mcr} shows a representative example of a \ac{mcr} \texttt{scenario} field from our GitGoodBench Lite. The \texttt{scenario} field contains the metadata based on which we compute the difficulty of this sample. In this case, the sample is hard, because there are multiple conflicts across multiple files. Furthermore, the sample contains the merge commit that serves as ground truth. We use the parent commits of this merge commit to generate a merge conflict that is resolved in the merge commit.

\subsubsection{\acf{fcc} Difficulty Heuristic}
\label{subsec:appendix:purity}
 For \ac{fcc} scenarios we define their difficulty through the purity of the \ac{fcc}:
\[
  d_{FCC}(p)=\begin{cases}
    \text{pure}, & 0.5 \leq p < 0.75\\
    \text{mixed}, & 0.75 \leq p < 1\\
    \text{noisy}, & \text{otherwise}
  \end{cases}
\]
where $p$ refers to the ratio of changes within the file for which we mined a \ac{fcc} to the overall changes in a \ac{fcc}. We consider any line with a Git change prefix (+ or -) a change. 

We expected this to capture the difficulty of \ac{ir} and \ac{icc} scenarios, because it captures how distributed changes the agent has to reason with are across files. The intuition being that it is easier to coherently generate commits and a plan for rebasing, when the reasoning spans fewer files. While, for \ac{fcc} samples easy scenarios exhibit the maximum \emph{success rate} of 100\%, they have a lower \emph{success rate} than medium and hard scenarios (\Cref{tab:appendix:total-success-rates}). One possible explanation could be that we are simply considering the ratio of changes and not the overall number of changes. A large overall number of changes forces the agent to reason across a much larger context window than a smaller number, yet in the purity-based difficulty heuristic we investigated, both are assigned the same difficulty.

\section{Prompts}
\label{appendix:section:prompts}
In this section we will provide the prompts used by our system for the individual scenarios and the LLM-as-a-Judge evaluation. For any missing details please refer to~\href{https://github.com/JetBrains-Research/git-good-bench}{our repository}.

\definecolor{promptorangeaccent}{RGB}{252, 128, 29}
\definecolor{promptorange}{RGB}{254, 204, 165}

\definecolor{promptlavender}{RGB}{230, 230, 250}
\definecolor{promptlavenderaccent}{RGB}{99,99,224}

\definecolor{promptturqoise}{RGB}{165, 235, 225}
\definecolor{promptturqoiseaccent}{RGB}{26, 121, 105}

\definecolor{promptpinkaccent}{RGB}{255,49,140}
\definecolor{promptpink}{RGB}{255,205,228}

\subsection{\acf{mcr} Scenarios}
\label{appendix:subsec:mcr-prompt}
In \Cref{fig:prompt:merge-conflict-resolution-1,fig:prompt:merge-conflict-resolution-2,fig:prompt:merge-conflict-resolution-3} we provide the prompt our system uses for \ac{mcr} scenarios. We show information on (1) the temporal ordering of the merge parent commits, (2) which conflicts occur (\texttt{git show} output) and (3) detailed instructions for resolving conflicts. Furthermore, we provide examples for the tools we provide in various conflict resolution contexts.
\begin{figure*}[t]
\begin{tcolorbox}[
  colback=promptlavender,
  colframe=promptlavenderaccent,
  title=\acf{mcr} Prompt - Part 1, width=\textwidth
]
\begin{small}
\begin{verbatim}
You are a staff software engineer with expertise in {programming_language} and git. 
    
You are helping a junior team member who has initiated a merge that resulted in 
one or more merge conflicts in one or more files. Your task is to help your
junior colleague with resolving all {total_amount_of_merge_conflicts} merge 
conflicts.

The semantic meaning and temporal relationship of the two sides of the merge 
conflicts are as follows for ALL merge conflicts you will encounter:
{commit_temporal_ordering}

The following files have merge conflicts:
{files_with_conflicts}

Below are all merge conflicts that need to be resolved, delimited by <CONFLICT-i> 
tags where i is the 0-based index:
{all_merge_conflicts}

\end{verbatim}
\end{small}
\end{tcolorbox}
\caption{Our \ac{mcr} prompt.}
\label{fig:prompt:merge-conflict-resolution-1}
\end{figure*}

\begin{figure*}[t]  
\begin{tcolorbox}[
  colback=promptlavender,
  colframe=promptlavenderaccent,
  title=\ac{mcr} Prompt - Part 2, width=\textwidth
]
\begin{small}
\begin{verbatim}
Instructions:
- Start with resolving the conflict at index 0 (CONFLICT-0) and proceed in 
ascending order through the conflicts.
    CONFLICT-0 is the current conflict that needs to be resolved.
- Consider the context around the merge conflicts, of the overall diffs and files 
in which the conflicts occur.
- Resolve the conflicts in a cohesive manner. For example, if you remove a function
in a conflict, make sure that you also remove any invocations of that function in 
any other conflicts.
- If you are just choosing one of the two sides, without changing any of the actual 
content, make sure to also reproduce the whitespaces exactly.
- If the merge conflict occurs due to a NOP (e.g. one side of the conflict is empty, 
the other is a commented code block) favor resolving the conflict to the most 
maintainable and concise way. Avoid dead code.
- Make sure to consider the implications your previous resolutions have on the 
remaining resolutions, especially when resolving multiple conflicts in a single 
file.
- If you find simple bugs, such as typos, copy and paste errors in variable 
assignments or parameters, feel free to help your junior developer fix these. 
Do not perform complex refactorings or attempt to change code drastically. 
Make as few changes to the side that you are accepting as possible.
- Consider the context of the temporal relationship of the branches that are being 
merged and the intent of the junior developer, with respect to which side of the 
conflict contains the local and which the incoming changes. The intent of the 
developer is to merge the incoming changes into the local changes.

You must only use the following tools and follow their specification exactly and 
always provide a reason for calling a tool.

All tools other than the ones listed below are INVALID and you MUST NOT use them 
under any circumstances.

Valid tools: 
- view_current_merge_conflict_with
- view_merge_conflict_at
- resolve_current_merge_conflict_with
- view_diff_for
- view_file_at: You must not use this command more than once per file as it is costly.

Below follow some examples detailing the usage of the above tools:
view_current_merge_conflict_with(context_window_size=15, reason='to get a more 
comprehensive overview of the local context around the current merge conflict')
view_current_merge_conflict_with(context_window_size=0, reason='to view only the 
current merge conflict without any local context')
view_current_merge_conflict_with(context_window_size=5, reason='to view only the 
current merge conflict with some local context')
view_merge_conflict_at(conflict_index=1, context_window_size=5, 
    reason='To ensure that the resolution for CONFLICT-0 is cohesive with 
    CONFLICT-1')
view_merge_conflict_at(conflict_index=1, context_window_size=10, 
    reason='To remind myself of the changes and context around CONFLICT-3 so 
    that I can decide whether to delete the import for ShoppingClient in the 
    current conflict')
view_diff_for(relative_path_from_project_root='src/app/io/FileParser.java', 
    reason='view the full diff between the local and incoming changes for the 
    file at path')
view_diff_for(relative_path_from_project_root='src/app/api/quantative_methods/
    regression.python', reason='understand how to resolve the current conflict such 
    that the resolution is cohesive and makes sense in the context of the overall 
    changes')
view_file_at(relative_path_from_project_root='src/tests/
    test_data_transformations.py', reason='understand the full context of the merge 
    conflict, because I think I might have found a small bug, but I need more context 
    to make sure it is one before applying a minor fix as part of the conflict 
    resolution.')
view_file_at(relative_path_from_project_root='src/app/utils.py', reason='to check 
    whether there are other commented out code blocks')
\end{verbatim}
\end{small}
\end{tcolorbox}
\caption{Our \ac{mcr} prompt continued.}
\label{fig:prompt:merge-conflict-resolution-2}
\end{figure*}

\begin{figure*}[t]  
\begin{tcolorbox}[
  colback=promptlavender,
  colframe=promptlavenderaccent,
  title=\ac{mcr} Prompt - Part 3, width=\textwidth
]
\begin{small}
\begin{verbatim}
view_file_at(relative_path_from_project_root='src/app/Authenticator.java', 
    reason='to see how the changes I made so far fit into the file and to ensure 
    I resolve the current conflicts such that it is cohesive with these previous 
    resolutions')
resolve_current_merge_conflict_with(content='from app.api.auth import 
    PremiumUser\\n', reason='The premium user class is a new authentication 
    class that is being used in the incoming changes and thus is most likely part 
    of what the junior developer wants to have access to')
resolve_current_merge_conflict_with(content='    bool debug = conf.shouldDebug;
    \\n    bool enableCaching = conf.enableCaching;\\n    bool shouldRetry = 
    conf.shouldRetry;\\n', reason='both of these configuration flags are being 
    used in the local changes, also I fixed a copy-paste bug and now the 
    enableCaching flag is correctly initialized to conf.enableCaching. The 
    shouldRetry flag is an incoming change that conflicts with what the developer 
    introduced, I will thus keep all three flags.')

Key Requirements:
- Once the conflicts are resolved you are done and must terminate.
- Your decisions must be cohesive across merge conflicts.
- Make sure that all your lines end with a newline character to avoid introducing 
accidental changes.
- You must always fill all parameters of the provided tools. This includes 
the "reason" parameter.
\end{verbatim}
\end{small}
\end{tcolorbox}
\caption{Our \ac{mcr} prompt continued.}
\label{fig:prompt:merge-conflict-resolution-3}
\end{figure*}

\subsection{\acf{ir} Scenarios}
\label{appendix:subsec:ir-prompt}
In \Cref{fig:prompt:interactive-rebase-1,fig:prompt:interactive-rebase-2,fig:prompt:interactive-rebase-3} we provide the prompt our system uses for \ac{ir} scenarios. We provide information on the commits participating in the rebase (\texttt{git show} output) to save agent turns spent reading the commit information. Then we provide detailed instructions for performing an interactive rebase. Finally, we provide examples for the tools we provide and the JSON schema the agent must use to interact with the rebase-todo file.
\begin{figure*}[!t]
\begin{tcolorbox}[
  colback=promptturqoise,
  colframe=promptturqoiseaccent,
  title=\acf{ir} Prompt - Part 1, width=\textwidth
]
\begin{small}
\begin{verbatim}
You are a staff software engineer with expertise in {programming_language} and Git.
You are helping a junior team member who has been committing all day without 
pushing their commits to the remote. Help them perform an interactive rebase 
to clean up their local tree. The rebase has already been initiated for you 
and is currently paused so that you can inspect the commits participating in 
the rebase and edit the rebase todo list.

The commits involved in the rebase are listed below. When referring to them in 
function calls, use the commit index "i" to refer to <COMMIT-i>. Avoid viewing 
all commits again, they are already presented below. The commits are delimited 
by the <COMMIT-i> and </COMMIT-i> tags:
{participating_commits}

Instructions:
Consider the changes in the commits and make adjustments if necessary such that the 
local tree:
- contains logically cohesive commits
- all commits have meaningful, descriptive commit messages that follow a unified 
format
- does not contain commits with duplicate commit messages
- follows best practices for maintainable code

You must only use the following tools and follow their specification exactly. 
Always provide a reason for calling a tool.
List of valid tools for this scenario: 
- view_rebase_todo: View current rebase todo list
- execute_rebase: Execute the rebase with the current rebase todo list, thereby 
all rebase-todo-list-items are processed in an ascending order
\end{verbatim}
\end{small}
\end{tcolorbox}
\caption{Our \ac{ir} prompt.}
\label{fig:prompt:interactive-rebase-1}
\end{figure*}

\begin{figure*}[!t]  
\begin{tcolorbox}[
  colback=promptturqoise,
  colframe=promptturqoiseaccent,
  title=\ac{ir} Prompt - Part 2, width=\textwidth
]
\begin{small}
\begin{verbatim}
- show_changes_in: If you want to spend more time thinking about some of the 
presented commits, use this tool to inspect the changes introduced by commit 
with index i
Below are some examples of how to use this function:
    show_changes_in(commit_index=4, reason='to inspect the changes in COMMIT-4')
    show_changes_in(commit_index=0, reason='to understand how the changes in 
    COMMIT-0 relate to its commit message')
- update_rebase_todo_list: Update the rebase todo list, reordering items or 
adjusting the commands to perform on commits. Each item in the list that you 
must pass to update_rebase_todo_list must be a string that complies with the
rebase-todo-list-item JSON schema specified below:
{{
    "type": "json",
    "schemaName": "rebase-todo-list-item",
    "schema": {{
        "type": "object",
        "properties": {{
                "commit_index": {{"type": "integer"}},
                "command": {{"enum": ["pick", "drop", "fixup", 
                            "fixup -c", "squash", "reword"]}},
                "commit_msg": {{"type": "string"}},
            }}
        }},
        "required": ["operations"],
        "additionalProperties": False
}}

Below are some examples of how to use this function:
Note: Positioning the rebase todo item with index 2 at the first position in 
the list, will swap it to the topmost position in the rebase todo list
    update_rebase_todo_list(rebase_todo_list_items=[
        '{{"commit_index": 2, "command": "pick"}}',
        '{{"commit_index": 1, "command": "reword", "commit_msg": "FIX: 
        Explicitly handle division by zero edge case"}}',
        '{{"commit_index": 0, "command": "fixup"}}',
        '{{"commit_index": 3, "command": "pick"}}',
        '{{"commit_index": 4, "command": "drop"}}'
    ], reason='to remove an unnecessary, noise, experimental commit, 
    improve the commit message of COMMIT-1 and consolidate the changes in 
    COMMIT-0 and COMMIT-1')
Note: Example for a different sample, you must ensure to always have 
exactly one item per commit.
    update_rebase_todo_list(rebase_todo_list_items=[
        '{{"commit_index": 0, "command": "pick"}}',
        '{{"commit_index": 2, "command": "squash", "commit_msg": "ADD: 
        Define interfaces and test cases for ShoppingBasketService"}}',
        '{{"commit_index": 1, "command": "pick"}}'
    ], reason='to reorder the local tree, yielding more coherent and logical 
    increments of changes in the local tree and to consolidate the changes in 
    COMMIT-0 and COMMIT-2')

Only the following commands are allowed for the rebase todo list items. Make 
sure to only provide the required fields for each command, all fields other than 
the required fields are invalid:
- pick: Use this commit as is. Required fields: ["commit_index", "command"]
- drop: Remove this commit. Required fields: ["commit_index", "command"]
- fixup: Meld this commit into previous commit, reducing the total amount of 
commits by 1. Only keep the previous commit's log message. Required fields: 
["commit_index", "command"]
- fixup -C: Meld this commit into previous commit, reducing the total amount of 
commits by 1. Only keep this commit's log message. Required fields: 
["commit_index", "command"]
- squash: Meld this commit into previous commit, reducing the total amount of 
commitsby 1. Commit message of resulting commit must be specified. Required 
fields: ["commit_index", "command", "commit_msg"]
- reword: Use commit, but edit commit message. Commit message must be specified. 
Required fields: ["commit_index", "command", "commit_msg"]
\end{verbatim}
\end{small}
\end{tcolorbox}
\caption{Our \ac{ir} prompt.}
\label{fig:prompt:interactive-rebase-2}
\end{figure*}

\begin{figure*}[!t]  
\begin{tcolorbox}[
  colback=promptturqoise,
  colframe=promptturqoiseaccent,
  title=\ac{ir} Prompt - Part 3, width=\textwidth
]
\begin{small}
\begin{verbatim}
Key Requirements:
- You must not simply pick all commits without modifying anything in the rebase 
todo list. Do your best to improve the local tree however you see fit.
- Avoid squashing all commits into a single commit, consider for which commits this 
would improve the resulting commit history.
- Try to consolidate the total size of the local tree such that the resulting tree 
has length k<{times_seen_consecutively}
- You must always fill all parameters of the provided tools. This includes the 
"reason" parameter.
\end{verbatim}
\end{small}
\end{tcolorbox}
\caption{Our \ac{ir} prompt.}
\label{fig:prompt:interactive-rebase-3}
\end{figure*}

\subsection{\acf{icc} Scenarios}
\label{appendix:subsec:icc-prompt}
In \Cref{fig:prompt:iterative-chunking-commits-1,fig:prompt:iterative-chunking-commits-2} we provide the prompt our system uses for \ac{icc} scenarios. First, we provide detailed instructions for chunking changes into logically cohesive commits that incrementally build toward the final patch. Next, we show the contents of the hunks the agent can select to save agent turns spent reading the commit information. Finally, we provide examples for the tools that the agent can use in these scenarios.
\begin{figure*}[!t]  
\begin{tcolorbox}[
  colback=promptorange,
  colframe=promptorangeaccent,
  title=\acf{icc} Prompt - Part 1, width=\textwidth
]
\begin{small}
\begin{verbatim}
You are a staff software engineer with expertise in {programming_language} and Git.
You are helping a junior team member who has been working all day without creating 
a commit to iteratively create commits and introduce their changes into the 
repository in a maintainable way. Help them to select hunks such that you can create 
multiple, small, but logically cohesive commits that are structurally sound, and 
follow best practices for maintainable code.

Instructions:
- Review the remaining hunks of code and help the junior engineer select the 
appropriate hunks for each commit.
- Ensure that you select as many hunks as you need to ensure structural integrity,
ie avoid breaking changes by, for example, removing a variable definition or 
initialization in one commit, but removing the usages of the variable in another 
commit.
- Identify the ids of the hunks that you should pass by the number following 
"HUNK-" in the list of remaining hunks below. For HUNK-8, the id you need to 
pass, if you want to select this hunk, would be 8.
- Each commit should be focused, small, and logically cohesive.
- Provide a clear and concise commit message for each commit following the format 
provided in the example usages.

Key Requirements:
- Avoid apply all changes in a single commit unless you are absolutely sure this 
will yield the best possible git history.
- You must always fill all parameters of the provided tools. This includes the 
"reason" parameter.

Process all of the following {number_of_remaining_hunks} hunks:
{remaining_hunks}

Task:
Pass a list of hunks to include in the commit and a descriptive commit message 
to the provided tool.

You must only use the following tools and follow their specification exactly 
and always provide a reason for calling a tool.
All tools other than the ones listed below are INVALID and you MUST NOT use them 
under any circumstances.
Valid tools:
- commit_changes_in
- commit_remaining_changes
\end{verbatim}
\end{small}
\end{tcolorbox}
\caption{Our \ac{icc} prompt.}
\label{fig:prompt:iterative-chunking-commits-1}
\end{figure*}

\begin{figure*}[!t]  
\begin{tcolorbox}[
  colback=promptorange,
  colframe=promptorangeaccent,
  title=\acf{icc} Prompt - Part 2, width=\textwidth
]
\begin{small}
\begin{verbatim}
Example usages:
    commit_changes_in(selected_hunks=[1,3], commit_message="FIX: Handle edge 
        case of uninitialized object",reason="to group the fixing of uninitialized 
        objects together")
    commit_changes_in(selected_hunks=[4], commit_message="ADD: Introduced 
        new enum class CarConfiguration", reason="to isolate the addition of the 
        new enum class")
    commit_changes_in(selected_hunks=[2,5], commit_message="REFACTOR: Migrate 
        car configurator to CarConfiguration enum", reason="The remaining changes 
        both deal with migrating the existing implementation to the enum introduced 
        in the previous commits. This way the commits build on each other in a 
        logical progression and the migration takes place once we ensure that the 
        class we migrate to is already present, thus avoiding breaking changes.")
    
    Once you have received a signal that you are done, you must always call 
    the tool in the example below to terminate:
    commit_remaining_changes(commit_message="UPDATE: Implement data 
        streaming feature", reason="because all hunks were processed and 
        I must now terminate")
\end{verbatim}
\end{small}
\end{tcolorbox}
\caption{Our \ac{icc} prompt continued.}
\label{fig:prompt:iterative-chunking-commits-2}
\end{figure*}

\subsection{LLM-as-a-Judge Evaluation}
\label{appendix:subsection:llm-eval}
In \Cref{fig:prompt:llm-eval-1,fig:prompt:llm-eval-2} we provide the prompt our system uses when evaluating the Git histories generated by the agent in \ac{fcc} samples. First, we provide detailed instructions regarding the dimensions based on which the \ac{llm} should assess the quality of a history. Next, we show the model one example response for each evaluation case. By doing so, we help the model follow the response schema. We also specify the response schema directly in the model configuration. Finally, we present the ground truth and agent-generated Git history. We use the same prompt for both evaluation runs when re-evaluating to mitigate the position bias.
\begin{figure*}[!t]  
\begin{tcolorbox}[
  colback=promptpink,
  colframe=promptpinkaccent,
  title=LLM-as-a-Judge Evaluation Prompt - Part 1, width=\textwidth
]
\begin{small}
\begin{verbatim}
Please act as an impartial judge and evaluate the quality of the two
git histories that are displayed below. Your evaluation should consider the 
following aspects:
- The quality of the commit messages with respect to consistency, conciseness, 
duplication and correctness with respect to the content of the commit.
- The logical cohesion of the changes present within the commits. Changes in 
a commit should have high logical cohesion.
- The logical progression and common thread between the commits and especially 
the order in which the commits are presented.
- The size of the commits. Commits should be as small as possible without 
breaking the system (e.g. changing a method signature in a non-backwards 
compatible way without also changing all uses of the method in the same commit).

Your job is to evaluate which git history is of higher quality. Avoid any position 
biases and ensure that the order in which the responses were presented does not 
influence your decision. Do not allow the length of the responses to 
influence your evaluation. Be as objective as possible. 

You must adhere to the response format demonstrated in example responses below:
{{
    'evaluation_result': 'HISTORY-1', 
    'evaluation_reason': 'The first git history has more descriptive commit and 
        non-duplicate messages that align much more accurately with the content 
        of the commits.' 
}}
{{
    'evaluation_result': 'HISTORY-2', 
    'evaluation_reason': 'The commits in git history 2 are more concise and 
        introduce logically coherent changes. The changes are introduces in such 
        a way that they are unlikely to break the system as the commits are self-
        contained with respect to the part of the system that they affect and 
        correctly propagate changes throughout the system. Thus I chose history 
        2 despite it having poorer quality commit messages.' 
}}
\end{verbatim}
\end{small}
\end{tcolorbox}
\caption{Our LLM-as-a-Judge evaluation prompt. We use the same prompt for both evaluation runs, we simply swap the positions of the histories that are evaluated in the prompt.}
\label{fig:prompt:llm-eval-1}
\end{figure*}

\begin{figure*}[!t]  
\begin{tcolorbox}[
  colback=promptpink,
  colframe=promptpinkaccent,
  title=LLM-as-a-Judge Evaluation Prompt- Part 2, width=\textwidth
]
\begin{small}
\begin{verbatim}
{{
    'evaluation_result': 'TIE', 
    'evaluation_reason': 'Both histories introduces changes that are logically 
        coherent and have similar commit messages. None of the two histories have 
        fundamental issues, such as duplicate commit messages or changes that
        obviously would break the system if they were introduced as presented. 
        As I am unsure, I am declaring a tie.' 
}}

<HISTORY-1>
{history_1}
</HISTORY-1>
<HISTORY-2>
{history_2}
</HISTORY-2>
\end{verbatim}
\end{small}
\end{tcolorbox}
\caption{Our LLM-as-a-Judge evaluation prompt continued.}
\label{fig:prompt:llm-eval-2}
\end{figure*}
\end{document}

%% file: acronym.tex
\begin{acronym}
\acro{llm}[LLM]{Large Language Model}
\acro{slm}[SLM]{Small Language Model}
\acro{cot}[CoT]{Chain-of-Thought}
\acro{swe}[SE]{Software Engineering}
\acro{vcs}[VCS]{Version Control System}
\acro{fcc}[FCC]{File-Commit Chain}
\acro{em}[EM]{Exact-Match}
\acro{ir}[IR]{Interactive Rebase}
\acro{icc}[ICC]{Iterative Committing of Changes}
\acro{mcr}[MCR]{Merge Conflict Resolution}
\acro{pl}[PL]{Programming Language}
\acro{mcp}[MCP]{Model-Context Protocol}
\end{acronym}